\documentclass{emulateapj}
\usepackage{subfigure}

\shortauthors{Straughn et al.}

\newcommand{\etal}	{\mbox{et al.\,}}

\newcommand{\Lo}	{\mbox{L$_{\odot}$}}
\newcommand{\cle}	{\ensuremath{\lesssim}}
\newcommand{\cge}	{\ensuremath{\gtrsim}}

\newcommand{\Hline}[1]{\mbox{H{\footnotesize {#1}}}}
\newcommand{\Ha}{\Hline{\mbox{$\alpha$}}\thinspace}
\newcommand{\Hbeta}{\Hline{\mbox{$\beta$}}\thinspace}

\newcommand{\OII}{[O\,{\sc ii}]}
\newcommand{\OIII}{[O\,{\sc iii}]}
\newcommand{\CIV}{C\,{\sc iv}}
\newcommand{\CIII}{C\,{\sc iii}]}

\begin{document}

\title{Emission-Line Galaxies from the PEARS Hubble Ultra Deep Field: A 2-D Detection Method and First Results}

\author{Amber N. Straughn\altaffilmark{1,2}, Gerhardt R. Meurer\altaffilmark{3}, Norbert Pirzkal\altaffilmark{4}, Seth H. Cohen\altaffilmark{5}, Sangeeta Malhotra\altaffilmark{5}, James Rhoads\altaffilmark{5}, Rogier A. Windhorst\altaffilmark{5}, Jonathan P. Gardner\altaffilmark{6}, Nimish P. Hathi\altaffilmark{2}, Chun Xu\altaffilmark{7}, Caryl Gronwall\altaffilmark{8}, Anton M. Koekemoer\altaffilmark{4}, Jeremy Walsh\altaffilmark{9}, Sperello di Serego Alighieri\altaffilmark{10}}

\altaffiltext{1}{amber.straughn@asu.edu}
\altaffiltext{2}{Department of Physics, Arizona State University, Tempe, AZ 85287}
\altaffiltext{3}{Department of Physics and Astronomy, Johns Hopkins University, Baltimore, MD 21218}
\altaffiltext{4}{Space Telescope Science Institute, Baltimore, MD 21218}
\altaffiltext{5}{School of Earth and Space Exploration, Arizona State University, Tempe, AZ 85287}
\altaffiltext{6}{Astrophysics Science Division, Observational Cosmology Laboratory, Goddard Space Flight Center, Code 665, Greenbelt, MD 20771}
\altaffiltext{7}{Shanghai Institute of Technical Physics, 200083 Shanghai, China}
\altaffiltext{8}{Department of Astronomy and Astrophysics, Pennsylvania State University, University Park, PA 16802}
\altaffiltext{9}{ESO Space Telescope European Co-ordinating Facility, D-85748 Garching bei Munchen, Germany}
\altaffiltext{10}{INAF - Osservatorio Astrofisico di Arcetri, I-50125 Firenze,
Italy}x


\begin{abstract}
The Hubble Space Telescope (HST) Advanced Camera for Surveys (ACS)
grism PEARS (Probing Evolution And Reionization Spectroscopically)
survey provides a large dataset of low-resolution spectra from
thousands of galaxies in the GOODS North and South fields.  One
important subset of objects in these data are emission-line galaxies
(ELGs), and we have investigated several different methods aimed at
systematically selecting these galaxies.  Here we present a new
methodology and results of a search for these ELGs in the PEARS
observations of the Hubble Ultra Deep Field (HUDF) using a 2D
detection method that utilizes the observation that many emission
lines originate from clumpy knots within galaxies.  This 2D
line-finding method proves to be useful in detecting emission lines
from compact knots within galaxies that might not otherwise be
detected using more traditional 1D line-finding techniques.  We find in
total 96 emission lines in the HUDF, originating from 81 distinct
``knots'' within 63 individual galaxies.
We find in general that \OIII\ emitters are the most common,
comprising 44\% of the sample, and on average have high equivalent
widths (70\% of \OIII\ emitters having rest-frame EW$>100${\AA}).
There are 12 galaxies with multiple emitting knots--with different knots exhibiting varying flux values, suggesting that the
differing star formation properties across a single galaxy can in
general be probed at redshifts $\cge0.2-0.4$.  The most prevalent
morphologies are large face-on spirals and clumpy interacting systems,
many being unique detections owing to the 2D method described here,
thus highlighting the strength of this technique.
\end{abstract}



\section{Introduction} \label{introduction}
It has long been known that galaxies display properties of their star
formation through emission lines, and because of this, systematic
studies of emission-line galaxies is an ongoing effort in order to investigate
galaxies' star formation--and thus evolution--throughout the history
of the universe.  Projects such as the KPNO International
Spectroscopic Survey (KISS; Salzer \etal 2000) have investigated
low-redshift emission-line galaxies' properties (Salzer \etal 2001 \&
2002).  Spectroscopic studies of faint, intermediate-to-high redshift emission
line galaxies have utilized large projects such as the CFRS (Lilly et
al. 1995, Hammer et al. 1997), COSMOS (Capak et al. 2007, Lilly et
al. 2007), and the DEEP1 and DEEP2 projects (Koo 1998, 2003; Willmer et
al. 2006; Kirby \etal 2007).  With the advantage of slitless grism
spectroscopy from the Hubble Space Telescope's (HST)
Advanced Camera for Surveys (ACS), larger samples of faint
objects --- reaching to $i'_{AB}\!\sim\!27.0$~mag --- are now possible.

Many detailed studies have arisen from projects such as these.
Earlier investigations have highlighted the importance of star
formation bursts in interacting galaxies in general (Larson \& Tinsley
1978), and subsequent studies have made use of emission-line fluxes to
arrive at star formation rates (SFRs; Kennicutt 1998).  In particular,
\Ha emission has been used to derive SFRs and those results have been
interpreted in the overall framework of galaxy evolution (Kennicutt
1983).  Several studies have highlighted the importance of
constraining the current SFR-density in the local universe using
emission-line galaxies (Gallego et al. 1995, 2002; Lilly \etal 1995), while others have
investigated the evolution of the SFR with redshift (Madau et
al. 1998; Cowie et al. 1999).  In the context of hierarchical merging,
active star formation has long been regarded as a strong indicator of
merging activity (Larson \& Tinsley 1978), and recent studies have
emphasized the evolutionary importance of merging galaxies and their
role in AGN growth over cosmic time, both theoretically (di Matteo et
al. 2005, Hopkins et al. 2005), as well as observationally (Straughn
et al. 2006, Cohen et al. 2006).  Studies of these types can be
greatly enhanced by larger samples of faint star forming or emission-
line galaxies at high redshift.

Slitless spectroscopy has been used often over the past several years
to detect emission-line galaxies.  In particular, HST's Near Infrared
Camera and Multi-Object Spectrometer (NICMOS) and Space Telescope
Imaging Spectrograph (STIS) instruments have produced several surveys
in which emission-line galaxies have been utilized to arrive at the
\Ha line luminosity function and SFRs (e.g., Yan et al. 1999; HST
NICMOS with the G141 grism), as well as the \OII\ luminosity function
and star formation densities at intermediate redshifts (e.g., Teplitz
et al. 2003; HST STIS with the G750L grism).  The ACS G800L grism has
also yielded very rich datasets, and the field of slitless
spectroscopy with HST has culminated the past few years with the HUDF
GRAPES (GRism ACS Program for Extragalactic Science; Pirzkal et
al. 2004, Malhotra \etal 2005) project, and more recently with the
PEARS (Probing Evolution And Reionization Spectroscopically) survey
(Malhotra \etal 2007, in prep., Cohen \etal 2007, in prep.),
which combined have yielded thousands of spectra over roughly half the
area of the GOODS North and South fields including the HUDF to
continuum fluxes of $i'_{AB}\!\lesssim\!27$~mag.  From the GRAPES
data, studies of emission-line galaxies have been performed and a
catalog has been compiled by Xu et al. (2007) using a 1D detection
method described briefly below.  Pirzkal et al. (2006) have performed
analysis of GRAPES emission-line galaxies' morphologies and evolution,
highlighting the importance of studying these objects at
$z\!\gtrsim\!1$.  A key advantage of this project over similar
ground-based studies is that the HST $i'$-band sky brightness is
$\sim\!3$ magnitudes darker than that from ground-based studies
(Windhorst et al. 1994).  With all the PEARS data analyzed, we
anticipate increasing the sample of faint emission-line objects by a
factor of 8--10 compared to the previous GRAPES project. In this
methods-oriented paper we describe in detail several techniques aimed
at detecting emission-line sources in the PEARS grism data and present
our data and results for emission-line galaxies detected in the HUDF
using a unique 2D line-finding method, which is shown to detect
roughly twice the number of sources as 1D methods on the same data.  A
subsequent paper will contain the complete catalog of emission-line
galaxies detected in the eight remaining PEARS fields, along with more
detailed analysis of their properties, including quantitative
morphological studies, star-formation rates, and line luminosity
functions.


\section{Data}
The PEARS ACS grism survey data consist of eight ACS fields with three
HST roll angles each (with limiting AB magnitude
$i'_{AB}\!\lesssim\!26.5$~mag; 20 HST orbits per field), plus the HUDF
field with four roll angles (limiting AB magnitude
$i'_{AB}\!\lesssim\!27.5$~mag; 40 HST orbits total) taken with the ACS
WFC G800L grism.  The G800L grism yields low-resolution
($R\!\sim\!100$) optical spectroscopy between 6000-9500{\AA}.  Four
PEARS fields were observed in the GOODS-N and five fields (including
the HUDF) in the GOODS-S.  A forthcoming data paper (Malhotra \etal
2007) will describe the PEARS project and data in detail.  A
description of the related GRAPES project can be found in Pirzkal
\etal (2004): both the PEARS and GRAPES projects contain grism
spectroscopy from the HUDF.  Roll angles for the PEARS HUDF are
$71^{\circ}$, $85^{\circ}$, $95^{\circ}$, and $200^{\circ}$.  This
paper will focus on emission-line galaxies detected in the HUDF, using
the optimal of several methods described in Section 3.  Preliminary
source extraction produced a large catalog of PEARS objects in all
nine fields with identifying numbers (Malhotra \etal 2007, in prep.).
These PEARS IDs will be used in this paper.


\section{Methods}
This paper focuses on our efforts at identifying an efficient and
robust method of detecting emission-line objects in the HST ACS PEARS
grism data, particularly in objects with knotty morphologies and in
continuum-dominated regions where lines might normally be missed.  To
this end, we have performed two separate, but related detections of
PEARS HUDF emission-line galaxies that both rely on a unique 2D
detection method, motivated by the observation that many emission
lines originate from clumpy knots within galaxies (Meurer et al. 2007,
Straughn et al. 2006b).  Our results from these two 2D methods
(hereafter ``2D-A'' and ``2D-B'', described in detail below) will be
compared to a separate method of detecting emission-line objects which
relies on searching for lines in 1D extracted spectra from the HUDF
PEARS data, as was also done for the GRAPES HUDF data (Xu et
al. 2007).

The 2D detection procedure begins with pre-processing of the grism
data, as described in detail in Meurer \etal (2007); here we give a
brief description.  Each image (both the grism and the direct
$i'$-band (F606W) image) was first ``sharpened'' by subtracting a 13x3
median smoothed version of the image from itself in order to remove
most of the continuum from the grism spectra, leaving mostly compact
features in the grism image.  In this step, the long axis of the
smoothing kernal is aligned with the dispersion axis of the grism.
These are primarily emission lines in individual object spectra, as
well as some residual image defects.  This method was
designed to detect lines in objects where continuum dominates and
lines would otherwise be washed out.  After this unsharp-masking
stage, the next step is to mask out the images of the zero-order in
the grism images.  This is accomplished by matching compact sources
found with the SExtractor program (Bertin \& Arnouts 1996) in both the
sharpened grism and direct images.  This defines a linear
transformation matrix which can be used to transform pixel coordinates
from the direct to the grism frame, as well as scaling factor between
the count rate in the direct image to that in the zero-order grism
image (as described by Meurer et al.\ 2007).  The geometric
transformation is also used to derive a precise calibration of the
row-offset between direct image sources and sources in first-order
spectra.  We determined that the transformation and row-offset are
stable with HST pointing and roll-angle, and hence adopted the same
values for all pointings.  The mask is made by using the count-rate
scaling to locate all the pixels in the direct image expected to be
brighter than the noise floor in their zero order grism image.  These
are transformed to the grism coordinates, grown in size by three
pixels to encompass the zero-order detection, and the resultant pixels
are set to zero.  Finally, SExtractor is used to catalog the masked
filtered grism images in order to arrive at a list of emission-line
source candidates, which is the input to both 2D emission-line source
selection methods.

\subsection{Method 2D-A: Cross-Correlation}
\begin{figure}
\includegraphics[scale=0.65]{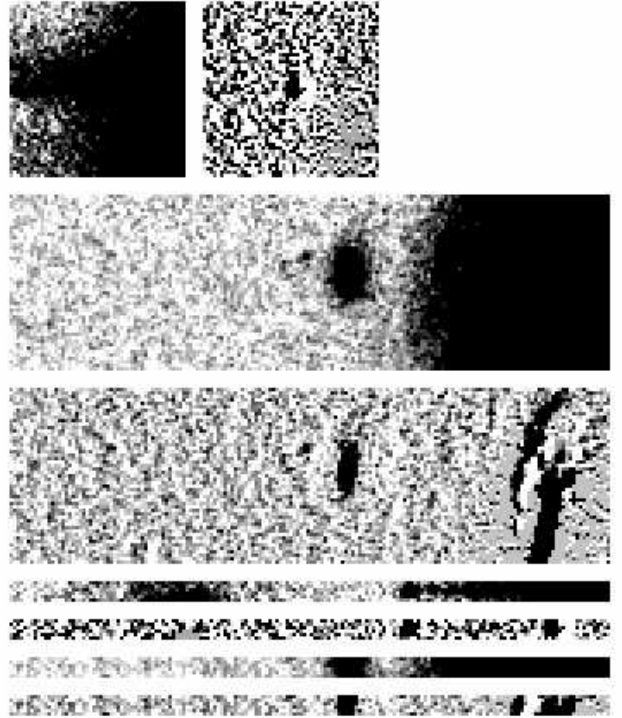}
\caption{Example of an object selected using Method 2D-A. Top panels show
unfiltered grism image spectrum (left) and the same spectrum after
median filtering (right).  Middle two panels show direct image source
both before and after the same median filtering.  Bottom four panels
show the 5-pixel wide ``ribbons'' (top two: grism; bottom two: direct)
used in the correlation step to determine which direct image source
the emission line originates from.}
\label{fig:2Dex}
\end{figure}

The first of the 2D methods (``2D-A'') is a blind selection that
relies on cross-correlation between the direct and grism sources, and
is partially interactive (Meurer \etal 2007).  Because of the
interactive step, it is desirable to limit the amount of known
contaminants that go into the algorithm.  Since stellar sources often
display a very strong continuum, sources with high SExtractor
elongation parameters (ELONGATION$>2.5$) in the dispersed grism image
were filtered from the catalogs to decrease the number of stellar
sources.  Sources that are very large or very small in the sharpened
grism image are filtered out by only selecting sources with SExtractor
parameter ``FWHM'' in the range of 1 to 10 pixels.  This filtering
reduces the number of sources that go into the 2D-A code approximately
by half.  Using these filtered catalogs, candidate emission lines are
examined first in the grism image, and then the corresponding direct
sources are located in the detection image.  This is a semi-automated
process in which lines in the grism image are displayed automatically,
and the validity of each source is subsequently determined by eye.
These potential emission-line sources are flagged as either: (1) a
star; (2) a grism- or detection-image blemish (in which two cases the
source is skipped); or (3) real, in which case the following is
performed.  For each ``real'' grism emission-line candidate, 5-pixel
wide ribbons are extracted from both the grism and direct images,
centered on the y-position (vertical in Fig.~\ref{fig:2Dex}) of the
source.  The grism image ribbon is then collapsed down to a 1D
spectrum.  This spectrum is then cross-correlated with the direct
image, and peaks are produced in the cross-correlation when knots
within the direct image are detected that correspond to the grism
image emission line.  Typically only one peak is found in the
cross-correlation yeilding a unique correspondence between line and
emitting source.  However, multiple peaks can occur due to the
presence of multiple knots within galaxies or separate galaxies in the
direct image ribbon.  In those cases the corresponding source is
selected manually.  The correct choice is usually obvious from the
location of the knot in the cross-dispersion direction (centered in
the ribbon) or from the shape of the knot compared to the emission
line in the filtered direct and grism images (Fig.~\ref{fig:2Dex};
also cf. Fig. 1 of Meurer et al. 2007).

The 2D-A line-finding software produces an output list for each
position angle with the detected emssion lines.  In many instances,
multiple knots with emission lines are detected in a single object.
Catalogs are then matched to determine which emission-line sources are
detected in multiple position angles.  The final catalog for the 2D-A
method was created by selecting sources which appear in at least two
position angles (PAs).


\subsection{Method 2D-B: Triangulation}
The second 2D technique (``2B-D") uses triangulation.  It starts with
the same input catalog as above, but without any prior filtering and
omission of sources based on their elongation and FWHM (however known
M stars are removed from the catalogs beforehand).  This method works
because each source, and hence emission line, was observed at more
than one PA on the sky, as is the case for our dataset.  The ACS grism
and ACS distortion are calibrated well enough so that one can map the
position of emission-line sources detected in a distortion corrected
grism image back onto the original distorted grism images, as well as
onto true sky coordinates of RA and Dec (instead of simply using the
detector x,y coordinates). When this is done for more than one PA, as
shown in Figure 2, one can infer the location of the source of the
emission line, which must necessarily lay somewhere along the
direction of the grism dispersion. Once the source of the emission
line has been inferred on the sky, we compute the wavelength of the
detected emission line independently and along all PA dispersion
directions (i.e. in all grism images where the line was detected).  A
true emission line source results in the same wavelength being derived
(within an error that we set to 40{\AA}, roughly one pixel), while a
spurious detection leads to inconsistent results where the computed
wavelength of a line is different when computed in different PAs.
Since we have more than 2 observations taken in more than 2 PAs for
this field, we actually used the method described above several times,
using different pairs of PAs (i.e. $71^{\circ}$ vs $85^{\circ}$,
$71^{\circ}$ vs 0$95^{\circ}$, etc..), as illustrated in Figure 2,
looking for emission-line sources that produce consistent results for
as many PA pairs as possible.

\begin{figure}
\includegraphics[scale=0.62]{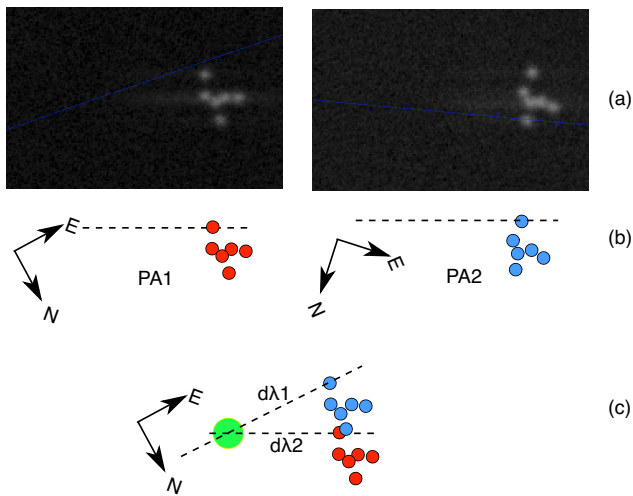}
\caption{The top two panels (a) show a series of emission lines
observed at two different position angles (PA1 on the left, PA2 on the
right). In both cases, the dispersion direction of the grism is nearly
horizontal, as shown in (b), where we noted the true direction of
North and East. Finally, the bottom diagram in (c) shows the remapping
of the grism dispersion (dashed lines) onto the true sky (e.g. RA and
Dec).  As shown in (c), once remapped onto the sky, the two dispersion
solutions intersect at a unique location (shown using the large
circle). The latter is the inferred location of the source of the
emission on the sky. We can then compute the expected wavelength of
the emission line following the PA1 dispersion relation (d$\lambda$1)
and following the PA2 dispersion relation PA2 (d$\lambda$2), and the two
should agree (within the expected uncertainty).}
\label{fig:traces}
\end{figure}

\subsection{Redshifts and line identifications}
Three separate catalogs were used to obtain redshifts for the selected
emission-line objects.  First, photometric and spectroscopic redshift
catalogs are from the GOODS-MUSIC sample (Grazian et al. 2006 and
references therein).  Spectro-photometric redshifts from Cohen \etal
(2007) were used to supplement the MUSIC catalog when no MUSIC
photometric redshift was available; this was the case for 16 objects.
We also use Bayesian photometric redshifts (BPZs) from Coe \etal
(2006) for the two objects (PEARS Objects 75753 \& 79283) that had no
spectroscopic redshifts, or MUSIC/Cohen photometric redshifts.
Sixteen sources have 2 emission lines, allowing an immediate redshift
determination using the ratio of line wavelengths which is invariant
with redshift (note that given the grism resolution of $R\!\sim\!100$,
\Hbeta and the \OIII\ doublet are usually blended).  About a third of the
sample ($\sim\!32$\%) has spectroscopic redshifts and 95\% have
photometric redshifts.  In total, three objects do not have any
published redshift.  Of these three, two (78237 Knot 1 \& 89209) have
2 lines each, and thus a redshift was determined based on the
wavelength ratios.  The redshifts are used to help identify the
emission lines in the grism spectra.  Spectra with two distinct lines
are in the minority (16 of 81 galaxy ``knots''); most spectra have a
single emission-line detection.  

For objects with only one emission line, line identification then
proceeds as follows.  For the given object's redshift (spectroscopic
when available; photometric otherwise), potential wavelengths are
calculated for \Ha, \OII, \OIII, Ly${\alpha}$, [MgII], \CIII, and \CIV.
An average spectro-photometric redshift error of
$<\!\delta\!z\!>\!=\!0.04\!\times\!(\!1\!+\!z\!)$ (Ryan \etal 2007,
Cohen \etal 2007, in prep., Coe \etal 2006) is used to calculate the likelihood
of an identification as follows.  Using the 4\% photometric redshift
error, a valid wavelength range for each potential line is calculated.
Here, we also include the estimated 20{\AA} wavelength calibration
uncertainty (Pirzkal et al. 2004) intrinsic to the grism data.  If the
detected emission line candidate falls within the calculated
wavelength range for any of the lines listed above, it is included in
our final catalog.  Once a confident line identification is made, we
use the wavelength to recalculate the redshift; these new redshifts
are given in Table 1.  In comparing grism redshifts to spectroscopic
redshifts, Meurer \etal (2007) arrive at a dispersion about unity of
0.007 for the 2D-detection method described here for secure line IDs
(i.e., sources with two lines, or \Ha or \OII\ emitters, as these are
typically the only plausible lines in the wavelength-range for that
redshift).  Line fluxes, rest-frame equivalent widths, and errors are
then calculated by fitting gaussian profiles to the spectra using a
non-linear least-squares fit to the given spectrum from five free
parameters: the gaussian amplitude, central line wavelength, gaussian
sigma, continuum flux level, and a linear term.  Here we include a
linear term in the fit to account for instances where the continuum is
not flat.  Line fluxes are averaged when the line is detected in two
or more roll angles.

\section{Results}
The primary goal of this work is arrival at a robust and efficient
technique to identify emission-line sources from the PEARS grism data
that is as automated as possible.  To this end, we have investigated
in detail two versions of a 2D detection method as described in the
previous section.  Methods 2D-A and 2D-B produced 75 and 96 lines
respectively, originating from multiple knots within galaxies.  This
is compared to 43 lines detected with the 1D method on the same data.
Details of the results of these comparisons are discussed here, as
well as a comparison to a catalog of emission-line galaxies generated
from the related GRAPES grism data (Xu \etal 2007) using the 1D
detection method.

\subsection{ELG detections from three different methods}
We summarize here in more detail detections resulting from three
different methods outlined above.  Note here the terminology used
resulting from our 2D method: ``lines'' (sources detected in the grism
image and hence distinct in position and wavelength), ``sources''
(which refer to individual clumps or knots within a galaxy), and
``galaxies'' (for example, one galaxy can contain three sources which
each have two lines).  The 1D line-finding method (as described in
detail by Xu et al. 2007 for the GRAPES data) involves selection of
emission lines from the 1D spectra generated by
aXe~\footnote{http:/stecf.org/instruments/ACSgrism/axe} with visual
confirmation.  For the PEARS HUDF, 62 candidate lines were detected,
19 of which were flagged with a quality code indicating a contaminant
or M-dwarf, resulting in a catalog of 43 PEARS galaxies.  These
remaining 43 galaxies are then compared to the catalogs generated by
the two versions of our 2D line-finding method.  Method 2D-A, the
cross-correlation technique (\S~3.1), produced a final catalog of 75
lines, all of which originate from valid faint emission-line sources,
since contaminants are thrown out in the user-interactive phase of the
process described above.  Individual PAs had 78, 114, 106, 77
detections in PAs $71^{\circ}$, $85^{\circ}$, $95^{\circ}$, and
$200^{\circ}$ respectively; 75 of these were detected in at least 2
PAs.  Method 2D-B, the triangulation technique (\S~3.2), produced a
total of 96 lines.  Method 2D-B also requires that an emission line be
in more than one PA; in the final sample of 96 lines obtained with
this Method, 12 were in two PAs, 38 were in three PAs, and 46 were in
all four PAs.

Method 2D-A, described in detail in the previous section, requires
some explanation of the results obtained since the software that
produces the catalog is partially user-interactive.  For PA085, two of
the authors (ANS and GRM) ran the blind emission line-finding software
on the data and compared results for completeness.  It was found that
there was a large (90\%) overlap in final sources obtained between
both users, suggesting that the method is robust in detecting secure
emission-line sources, and user dependancies introduce relatively
little bias.  An investigation of the sources that were selected by
one user and not the other shows several cases of multiple emission
lines in knotty galaxies that often were offset from the other user's
detection by only a very small amount.  There are a few cases of
isolated galaxies where only one user detects a line.  In general, the
differences appear to be legitimate operator differences, and account
for $<\!10$\% of the detections.  Those operator errors that are false
detections are effectively weeded out by the requirement that the
sources match in multiple PAs.

As described above, Method 2D-B (which uses traces in the grism image
to determine the direct image emitting source) detects the most real lines
in the PEARS data and discards spurious detections automatically.
Because of this, it appears to be the most efficient and robust
technique to detect emission-line sources in the grism data, and below we
compare this method to the other two methods described here (Method 1D
and Method 2D-A).

\subsection{Comparison of Method 2D-B to Method 1D}
\begin{figure}
\includegraphics[scale=0.5]{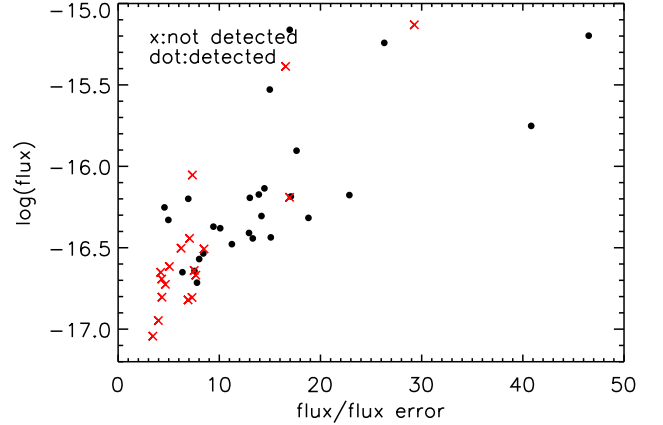}
\caption{Signal to noise estimates of PEARS objects detected with the
1D Method.  Black dots are 1D-detected objects that are also detected
with our 2D-B Method; red x's are 1D-detected objects that are missed
by the 2D-B Method.  Fluxes are from Xu \etal.  In general, objects
missed by Method 2D-B are lower S/N, clustered below S/N$\sim\!8\!-\!10$.}
\label{fig:s/n}
\end{figure}

First, we compare to the 1D method used on the same PEARS data.  Here
we find that Method 2D-B detects $1.9\!\times$ as many sources for
this field.  Overall, the overlap between the two methods is 72\%,
with the 1D method detecting 12 unique sources and the 2D method
detecting 50 unique sources.  We note here that the input to the 1D
method is the SExtractor catalog of entire galaxies, in contrast to
our SExtractor catalog of individual galaxy knots.  Thus in regard to
the 1D method, the emission lines are diluted by the continuum and the
equivalent width goes down below the detection limit.  An inspection
of the initial 2D-generated input files in comparison to these 12
1D-detected but 2D-undetected sources shows several aspects of
interest.  First, the majority of these 12 1D-detected sources are
clustered along the edges of the field, with only 3 of them extending
inwards more than 700 pixels (or 1/5 the width of the image).  This
suggests---and was confirmed on more detailed inspection---that many
times the object in question is undetected in at least one PA
(sometimes up to three PAs), and would thus not make it into the final
catalog produced by the 2D-B method.  Second, we notice that only in a
very few cases are there any SExtractor detections located along the
dispersion direction for any given 1D-only detected object.  This
shows that these sources were not in the input SExtractor catalogs
because they were below our 2D-detection limit, thus explaining the
absence from our final 2D-B ELG catalog.  Figure~\ref{fig:s/n} shows
an estimate of signal-to-noise for these 1D-detected objects, and
indicates that the 1D-detected objects that were missed by the 2D
method were in general lower S/N and likely below our detection limit
imposed in the initial grism emission-line catalog selection.

\begin{figure}
\plotone{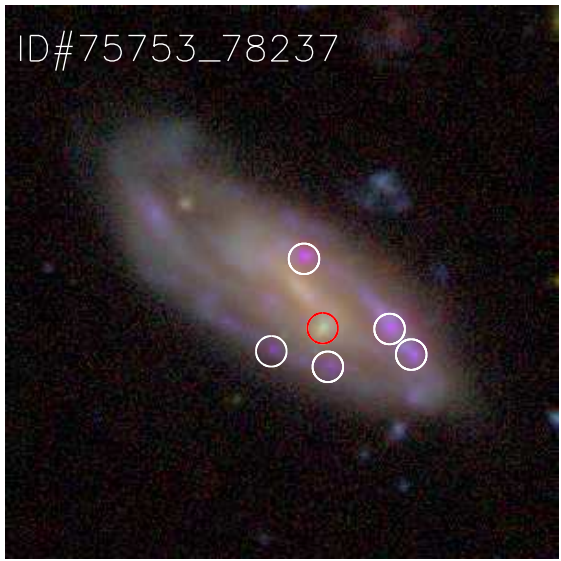}
\caption{PEARS Objects 75753 \& 78237.  Stamp is 9'' across; this
galaxy has a redshift $z=0.339$.  These were extracted as two sources
but are part of the same galaxy.  The knot indicated by the red circle is
likely an interloper with an undetermined redshift.  The other five
knots all contain \Ha and/or \OIII\ emission; flux values are given in
Table 1.  This composite image (as well as Figures 5 \& 6) was constructed using the HUDF $B, V, i', z'$ data.}
\label{fig:75753}
\end{figure}

\begin{figure}
\plotone{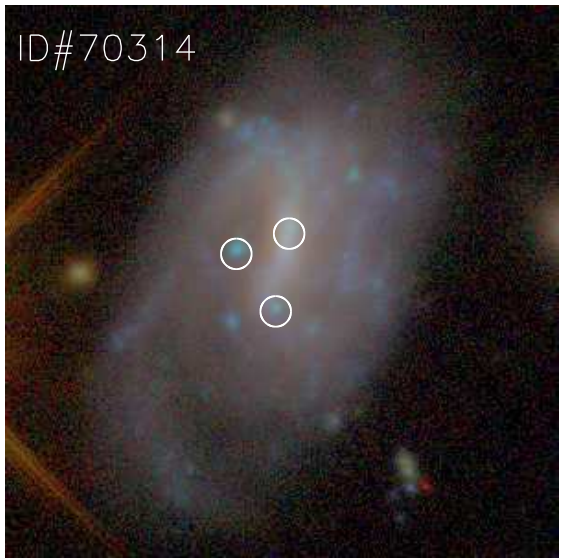}
\caption{PEARS Object 70314.  Stamp is 9'' across; this galaxy has a redshift $z=0.144$.  All three knots have \Ha emission.
The line from the nuclear region of the galaxy has an equivalent width
$\sim$4$\times$ smaller than that from the other two knots.  This result
highlights the strength of the 2-D Method utilized here: this galaxy
has no detected lines using the 1-D Method in either the PEARS or GRAPES
data, presumably because the line flux was overwhelmed by continuum
flux when light from the entire galaxy was extracted.  However,
narrowing in on individual knots allows us to see the line emission.}
\label{fig:70314}
\end{figure}

\begin{figure}
\plotone{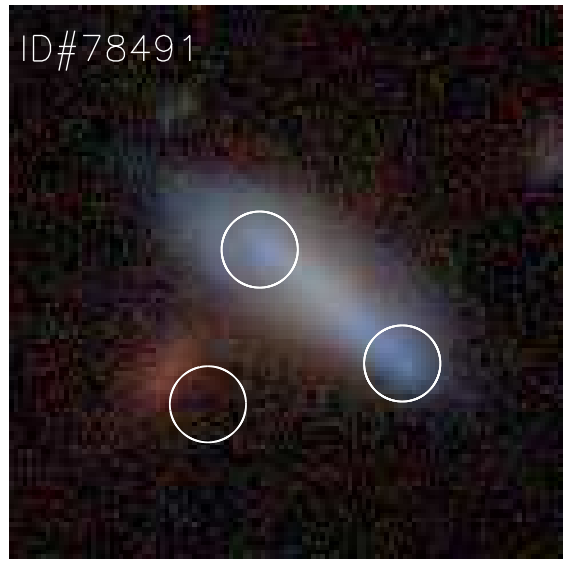}
\caption{PEARS Object 78491.  Stamp is 3.6'' across; this galaxy has a
redshift $z=0.234$.  The two blue knots on the ends of the galaxy each
have \Ha and \OIII\ emission; the knot on the left-hand side of the
galaxy has stronger \OIII\ flux by a factor of $\sim\!2$ and the
right-hand side knot has slightly stronger \Ha flux.  The third
circled ``knot'' is clearly not part of this PEARS galaxy, although it
is emitting a very strong line ($EW\!=\!237.0$\AA) at
$\lambda\!=\!7143$\AA.  No redshift is available for this object,
however, so line identification is not possible.}
\label{fig:78491}
\end{figure}

Figures~\ref{fig:75753}-~\ref{fig:78491} show some examples of
2D-detected galaxies with several emitting knots.  These composite images were created using the HUDF F435W ($B$), F606W ($V$), F775W($i'$), and F850LP ($z'$) data.  Objects 70314 and
78491 were not detected using the 1D technique.  This is likely due to
continuum flux dominating the spectrum, an effect which was mitigated
in our technique by the sharpening process (see Section 4.6 for a full
discussion of these objects).  Therefore, in general, it is shown that
the overlap between Method 1D and Method 2D-B is large, and the 1D
objects missed by the 2D-B Method are due to our imposed detection
threshhold.  In total, the 2D Method finds almost twice as many
sources.

\subsection{Comparison of Method 2D-B to Method 2D-A}
Given that Method 2D-B was developed in conjunction with Method 2D-A,
a comparison between these two methods is warranted as well.  As
described above, both methods have identical input catalogs for each
PA, with the exception that the input to method 2D-A was pared down to
avoid selection of undesired objects (i.e. stars) in the
user-interaction phase of the analysis.  Comparison of Method 2D-B to
Method 2D-A shows that all but 2 sources detected by Method 2D-A were
also detected by Method 2D-B, an overlap of 96\%.  Additionally, the
level of human interaction is greatly reduced in Method 2D-B, making
it both more efficient and reliable.

\subsection{Comparison of Method 2D-B to GRAPES catalog}
\begin{figure}
\includegraphics[scale=0.48]{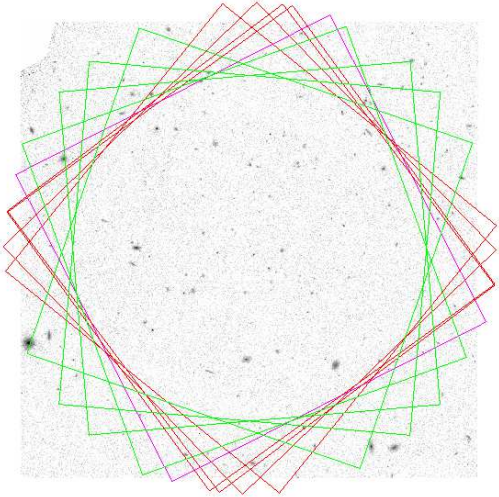}
\caption{The four PEARS HUDF pointings are shown in green, as well as
the four GRAPES shown in red (plus one archival ACS grism field uses in the GRAPES
study shown in magenta; see Sec. 2).  All PEARS and GRAPES fields are
centered on the HUDF.  Eight additional PEARS fields will be analyzed
in a future study: four fields in the GOODS-N and four more in the
GOODS-S.}
\label{fig:fields}
\end{figure}

Although the GRAPES project (Pirzkal \etal 2004, Malhotra \etal 2005)
involves a different dataset than the one used for the present work, a
comparison of our results to the previous GRAPES ELG catalog in Xu
\etal (2007) is explored here since the data are for approximately the
same field.  Figure~\ref{fig:fields} gives a graphical comparison of the PEARS and GRAPES fields centered on the HUDF.  The process used to arrive at the GRAPES ELG catalog is
the same as ``Method 1D'' described above, with some manual additions
(approximately 10\%) of objects after visual examination of all the
individual spectra as described in that paper.  The first difference in
the two datasets is that the Xu \etal (2007) ELG catalog utilized the
GRAPES data (40 HST orbits) plus one epoch of preexisting ACS grism
HUDF data, increasing the observed grism exposure time by about
one-fifth (Pirzkal \etal 2004) and also increasing the overall
combined area observed.  Second, since the fields are not exactly
overlapping, there are some GRAPES ELG objects that are not in the
PEARS fields and vice versa.  Given these factors, the comparison is
not as straightforward as, e.g., the comparison to the 1D Method
applied to the identical PEARS data, as described above.  However,
when doing the comparison, we find that 61\% of the 2D-B detected
sources are in the GRAPES catalog, with 37 unique lines appearing in
the 2D-B catalog only.  Of the 39\% of GRAPES sources unique to the GRAPES
catalog, many are found to exist outside of the PEARS observing area.
Specifically, 44\%, 34\%, 34\%, and 27\% of the 2D-B-undetected GRAPES
sources fall outside the four PEARS roll angles $71^{\circ}$,
$85^{\circ}$, $95^{\circ}$, and $200^{\circ}$ respectively.  In total,
there are 35, 41, 41, 45 sources in the four respective PEARS PAs
that are not detected with Method 2D-B.  The sources that were
detected with the 1D Method from GRAPES but were missed by 2D-B in
general were missed for the same reason as with Method 1D used on the
PEARS data (as described in Sec. 4.2): those missed were below our S/N
threshhold required for Method 2D-B (Fig.~\ref{fig:s/n}).  In particular, the
missed objects generally have S/N$\sim\!2\!-\!3$, while most of our objects
detected in GRAPES generally have higher S/N values.  We thus conclude
that this is the same effect as was seen when comparing to Method 1D
for the PEARS data.  This is expected given our detection limit which
serves to greatly increase the reliability of our 2D detection method.

\subsection{ELG catalog and statistics}
\begin{figure}
\subfigure[1D Extraction]{
\includegraphics[scale=0.141]{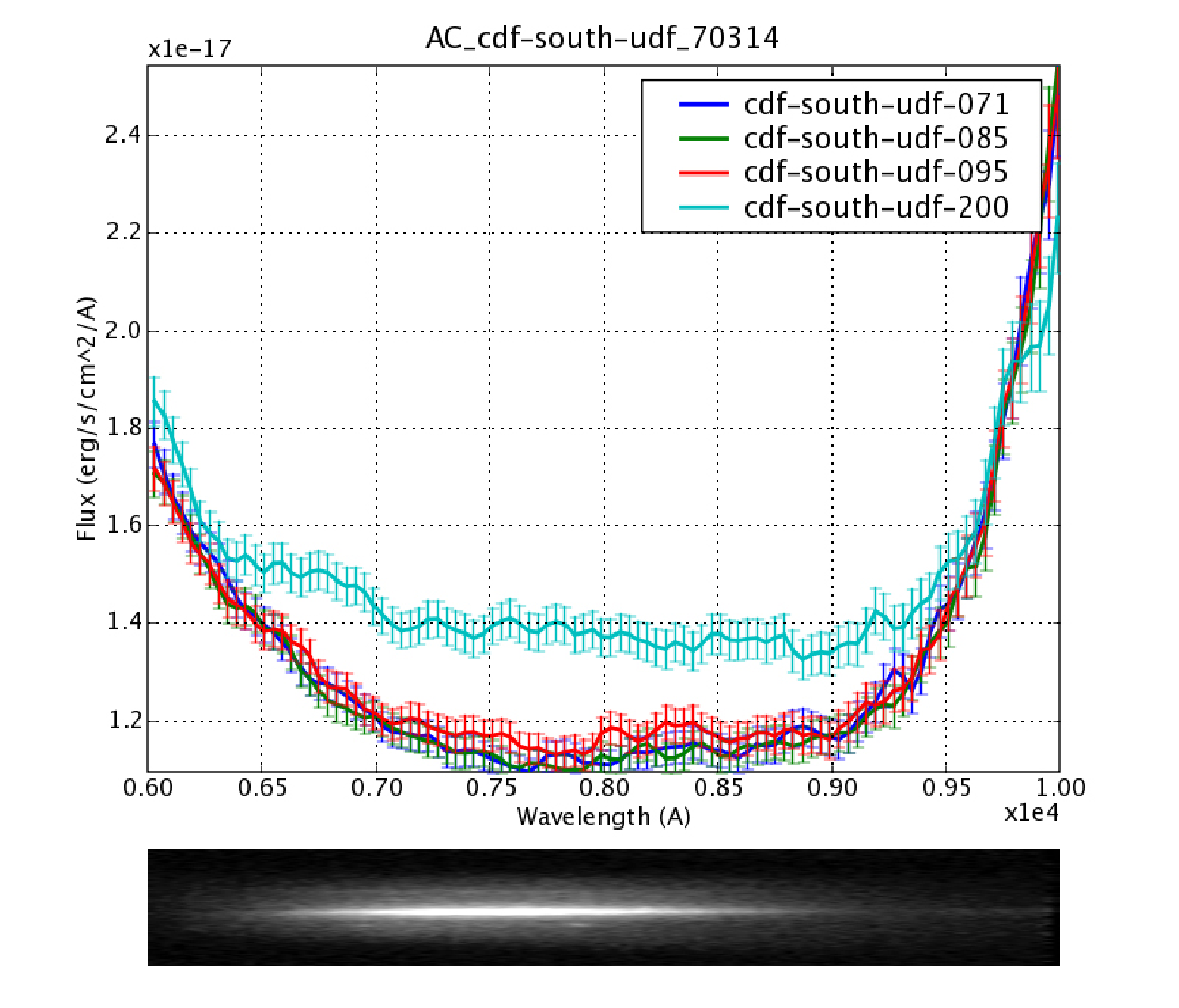}}
\hspace{0cm}
\subfigure[2D Extraction]{
\includegraphics[scale=0.151]{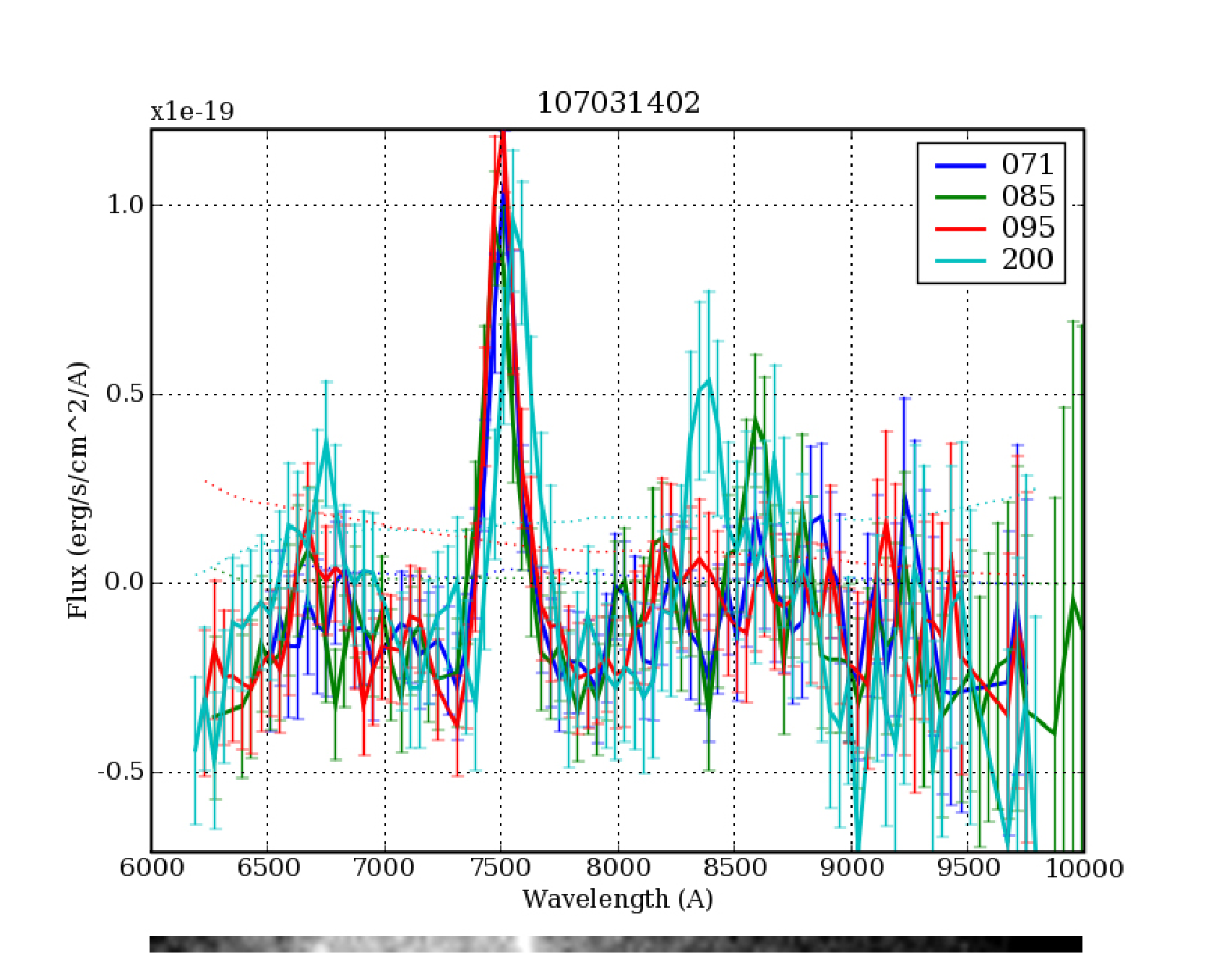}}
\caption{(a): 1D extraction of PEARS Object 70314 (entire galaxy; see
also Figure 12).  Top panel shows the 1D spectrum from 1D extraction;
bottom panel shows the 2D spectrum of this object.  Top-right inset
gives labels for the four HST roll angles used in this dataset
($71^{\circ}$, $85^{\circ}$, $95^{\circ}$, and $200^{\circ}$).  Scale
of y-axis is in units of $10^{-17}$ (b): 2D extraction of Knot \# 2 in
the same galaxy.  Panels and inset are the same as in (a).  Here the
y-axis scales in $10^{-19}$.  This is an example of the success of the
2D extraction method: no lines are detected when the 1D extraction of
the entire galaxy is performed.  However, the 2D extraction of three
separate emitting knots in this galaxy reveals strong emission lines
in the spectra of all three knots.}
\label{fig:exspec}
\end{figure}

\begin{figure}
\plotone{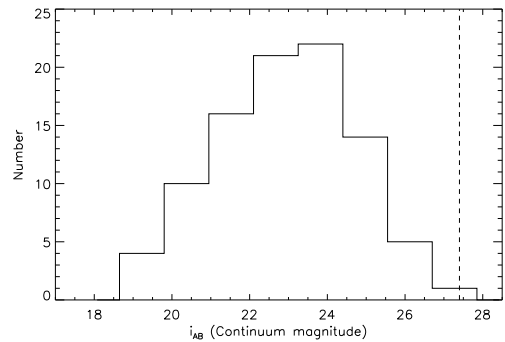}
\caption{Histogram of $i'$-band magnitudes of all PEARS HUDF
emission-line objects, showing a peak in the distribution around
$i'_{AB}\!\simeq\!24$ mag.  The PEARS HUDF continuum detection limit
is $i'_{AB}\!=\!27.4$ mag (Malhotra \etal 2007, in prep.).}
\label{fig:maghist}
\end{figure}

Our final catalog of PEARS HUDF emission-line objects, derived from
the most efficient method investigated here---Method 2D-B---is given
in Table 1.  In total, 96 distinct lines were detected in 81 galaxy
sources or ``knots'' in 63 PEARS galaxies.  Examples of galaxies with
several emitting knots are shown in
Figures~\ref{fig:75753}--~\ref{fig:78491}, demonstrating the strength
of the 2D Method as compared to the 1D Method.  In addition,
Figure~\ref{fig:exspec} shows how the 2D Method is able to detect
lines in galaxies that were undetected by the 1D Method due to strong
continuum overwhelming the emission lines.  The percentages of
identified lines are as follows: 34\% are \Ha, 14\% are \OII, and 44\%
are \OIII, with 3\% accounting for other less common lines ( MgII,
\CIII, and \CIV).  Our catalog includes 39 new spectroscopic redshifts
for galaxies that are on average fainter than the standard magnitude
limited redshift survey ($z'_{AB}\!=\!23.5$~mag for ground-based GOODS
spectroscopic redshifts; Elbaz \etal 2007).  The faintest ELG has a
continuum $i'_{AB}\!=\!27.4$~mag, and the average continuum magnitudes
of \Ha, \OII, and \OIII\ emitting galaxies are $i'_{AB}\!=\!21.9$,
$24.1$, and $23.6$~mag respectively.  The magnitude distribution of
the sample is given in Figure~\ref{fig:maghist}.

\begin{figure}
\plotone{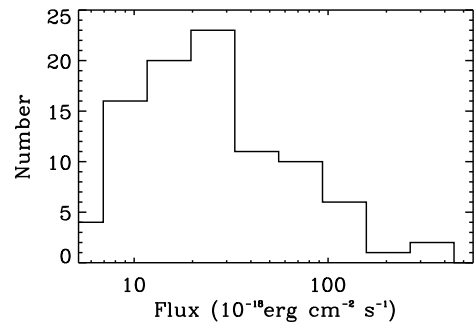}
\caption{Line flux distribution of all lines detected with Method 2D-B, in units of $10^{-18} ergs cm^{-2} s^{-1}$.}
\label{fig:fluxdistall}
\end{figure}

\begin{figure}
\plotone{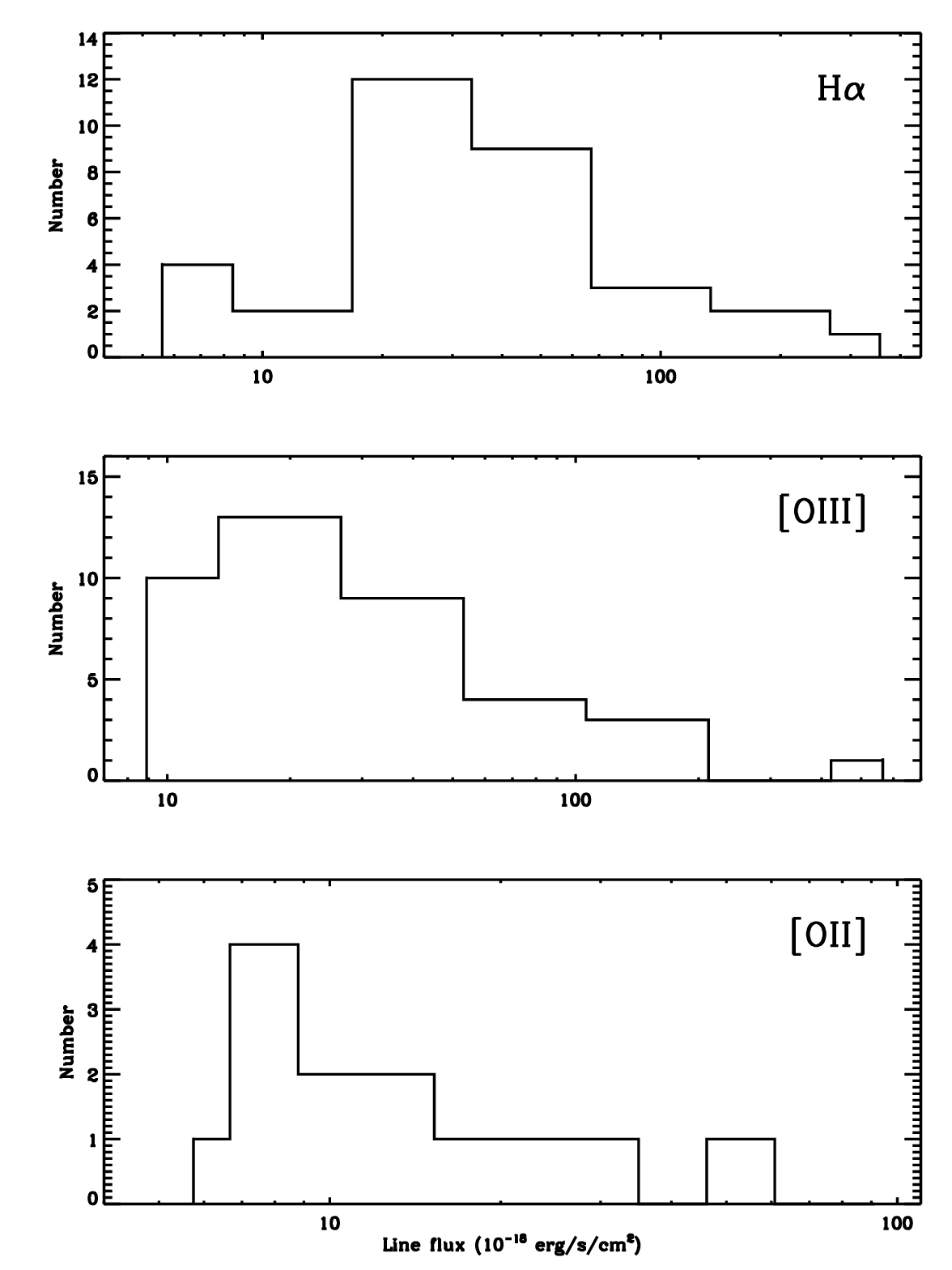}
\caption{Individual flux distributions of \Ha, \OIII, and \OII\ lines.  The distributions peak at $\sim$$2.5\times10^{-17}$ ergs cm$^{-2}$s$^{-1}$ for \Ha and \OIII, and near $5.0\times10^{-18}$ ergs cm$^{-2}$s$^{-1}$ for \OII.}
\label{fig:fluxdistindiv}
\end{figure}

The faintest line flux is $5.0\!\times\!10^{-18}$~ergs
cm$^{-2}$s$^{-1}$, with the average line flux being
$3.9\!\times\!10^{-17}$~ergs cm$^{-2}$s$^{-1}$.  The \OIII\ emitters
have on average high equivalent widths, with 70\% of them having
$EW\!>\!100${\AA}.  The line flux distribution for all sources is
given in Figure~\ref{fig:fluxdistall}, while
Figure~\ref{fig:fluxdistindiv} gives the line flux distributions for
the individual lines.

\begin{figure}
\plotone{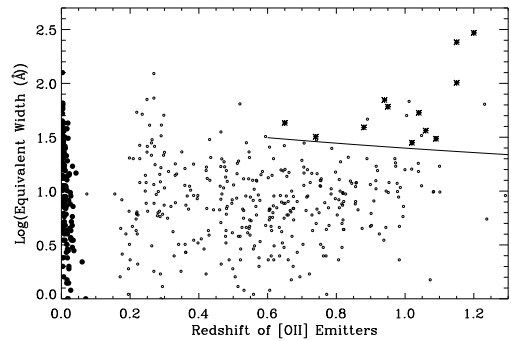}
\caption{Rest-frame equivalent width as a function of redshift for
[OII] emitters.  Stars are PEARS HUDF \OII\ emitters, filled dots are
from Jansen \etal (2000) and small dots are from the Canada-France
Redshift Survey (Hammer \etal 1997).  For our sample, \OII\ can be
detected in the redshift range $0.6\!\lesssim\!z\!\lesssim\!1.5$ given
the grism range 6000-9500{\AA}.  Approximate PEARS selection limit is
given by solid line.}
\label{fig:zvsew}
\end{figure}

An interesting potential trend appears in Figure~\ref{fig:zvsew},
which shows the equivalent width of PEARS HUDF \OII\ lines as a
function of redshift as compared to nearby galaxies from Jansen \etal
(2000) and intermediate redshift galaxies from the CFRS sample (Hammer
\etal 1997).  It is clear that the EW of the PEARS \OII\ sources are
extremely high compared to local samples--especially above $z\sim1.1$.
Here we note that grism selection of \OII\ emitting regions
systematically selects higher-EW objects in the \OII\ redshift range
probed by the grism (which could be due to the smaller HST PSF
including less continuum from the surrounding area of the knot, thus
raising the observed EW; see the sensitivity limit plotted in
Fig.~\ref{fig:zvsew}).  Hence the fact that the average EW is higher
than local galaxies is not surprising.  However, the discovery of
\OII\ emitters with EW$>100$\AA\ is interesting.  These are
exceedingly rare in the local universe (Jansen \etal 2000), and here
we only see them at the highest redshifts ($z>1.1$).  As shown in
Fig.~\ref{fig:zvsew} they are also known from previous ground-based
surveys (CFRS; Hammer \etal 1997), and appear to be more common with
increasing $z$ (Cowie \etal 1996).  Sources with similar (and higher)
EW(\OII) were also reported in previously published HST slitless
observations (Meurer \etal 2007; Teplitz \etal 2003).  Teplitz et al.
(2003), in particular, find a high incidence of EW \cge 100\AA\
\OII\ emitters at z\cle 0.5, while noting that at high redshifts, lower
EW lines might have been missed in comparative surveys.  While
statistics are low presently, and thus no definite statement can be
made concerning this trend, several possible explanations of its origin exist.
For example, strong evolution of galaxies' star formation properties
with redshift would cause this occurance of very high \OII\ EW at high
redshift.  Additionally, lower extinction between the sources of
ionizing radiation and the gas would produce higher EW values.
However, this phenomenon could also be caused by cosmic variance; the
comoving volume of the PEARS HUDF is clearly much less than the local
sample and the 0.2 \cle z \cle 1.0 sample, and this effect could influence
results presented here.  As an example of this effect, Takahashi \etal
(2007) found that \OII\ emitting star-forming galaxies in the COSMOS
field show clustering tendencies at redshifts z\cge 1.2.  We
anticipate a better study of this phenomenon when the other eight
PEARS fields are analyzed and simulations of the data are available to
aid in sorting out various selection effects.

\begin{figure}
\plotone{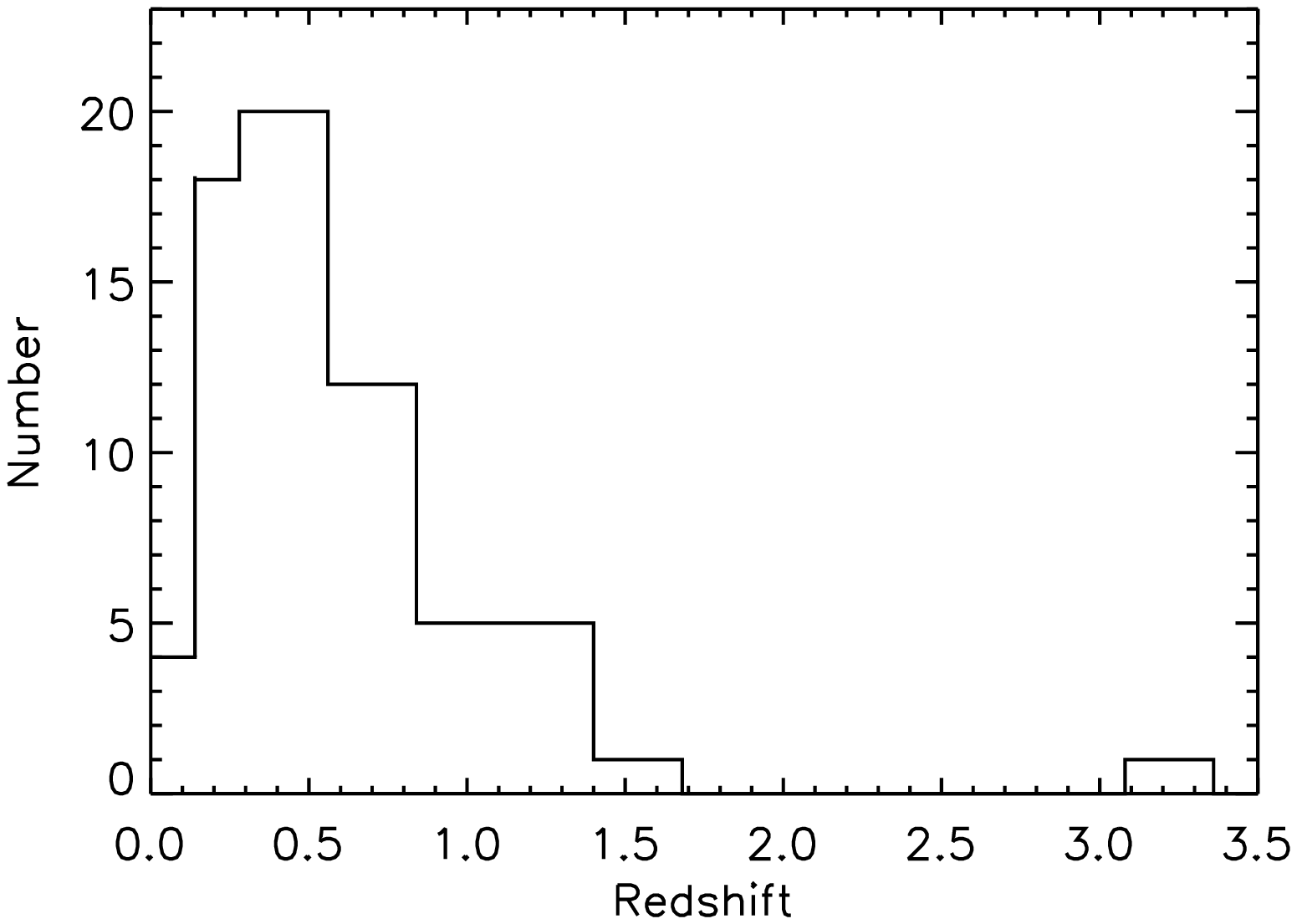}
\caption{Redshift distribution for PEARS ELGs showing peak of
distribution around z$\sim$0.5.  Given the grism properties (sensitive
from 6000{\AA} to 9500{\AA}), the \Ha line is observable from
0$\cle$z$\cle$0.4; \OIII\ from 0.1$\cle$z$\cle$1.1, and \OII\ from
0.4$\cle$z$\cle$1.5.  The one object at higher redshift in this figure
is the AGN (\CIV\, \CIII\ emitter) at $z=3.17$.  Because the identified
lines are only available at these particular redshifts (and are
generally the strongest lines in star-forming galaxies), the
emission-line N(z) peaks at a lower $z_{max}$ than the field galaxy
photometric redshift distribution which peaks at $z\sim1-1.5$ (Ryan
\etal 2007, Cohen \etal 2007, in prep.).}
\label{fig:zdist}
\end{figure}

The redshift distribution of these ELGs is shown in
Fig.~\ref{fig:zdist}, with the majority of redshifts lying between
$z\!=\!0$ and $z\!=\!1.5$ and the peak occuring around $z\!\sim\!0.5$.
Since the identified lines---which are only observable at the
redshifts in the plot---are generally the strongest lines in
star-forming galaxies, this explains why the emission-line N(z) peaks
at a lower $z_{max}$ than the field galaxy photometric redshift
distribution which peaks at $z\!\sim\!1\!-\!1.5$.  This is thus in
part an artifact of the ACS grism selection function (see Malhotra
\etal 2005).

A qualitative look at the morphologies of the emission-line galaxies (in $B,V,i',z'$; see Fig. 4-6)
suggests that the majority of these objects are clumpy, knotty
galaxies that have distinct emitting regions of presumably active star
formation.  In particular, we find many face-on knotty spirals, as
well as clumpy interacting systems with regions of enhanced star
formation that were missed with the 1D Method.  A subsequent paper
will investigate the emission-line galaxies' morphologies in a
quantitative manner, including the results of the selection for the
entire PEARS dataset in addition to these HUDF ELGs.

\subsection{Line luminosities of PEARS galaxies: Comparison to nearby galaxies}
\begin{figure}
\plotone{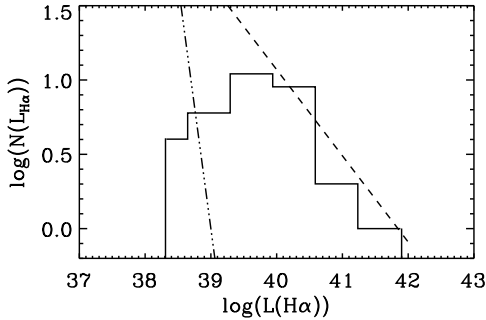}
\caption{Distribution of \Ha line luminosities of the PEARS galaxies
(median redshift of z$\sim$0.26), with the local Kennicutt \etal
(1989) sample's bright-end slope plotted as a power-law (dot-dashed
line; a=-3.3).  The PEARS sample slope differs from the local one (a=-0.58); we
detect more bright \Ha emitting regions in the grism data. }
\label{fig:halum}
\end{figure}

From our sample, there are 33 \Ha and 13 \OII\ emission regions from
galaxies at average redshifts of $z\!\sim\!0.26$ and $z\!\sim\!1.05$
respectively.  Here we discuss the properties of these objects in
terms of their line luminosities in comparison to local samples from
Kennicutt \etal (1989) and Zaritsky \etal (1994).
Figure~\ref{fig:halum} gives the luminosity histogram of PEARS HUDF
\Ha emitters (solid line) with the slope of the local HII region \Ha
distribution from Kennicutt \etal (1989) as a dot-dashed line.  We
have also plotted the best fit line of the bright end of our distribution with
a dashed line.  Here we see that the grism observations do not detect
some of the fainter emission, as is expected; however, we do detect
brighter sources, lending to the shallower slope.  We note here that
this effect is not due to spatial resolution: at $z\!\sim\!0.3$, each pixel is
$\!\sim\!130$ parsecs, and Kennicutt \etal (1989) show that there is
almost no difference in the luminosity histograms when degrading the
spatial resolution from 30 to 300 parsecs.

\begin{figure}
\plotone{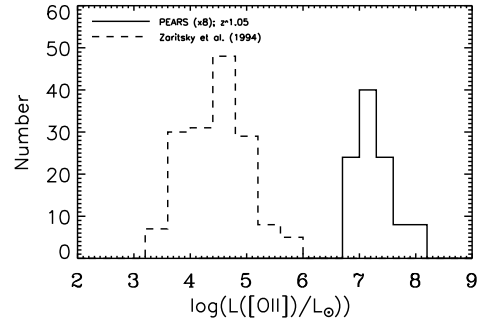}
\caption{Distribution of \OII\ line luminosities of the PEARS galaxies
(solid line multiplied by a factor of 8; median redshift of
$<\!z\!>\!\sim\!1.05$) and of HII regions within nearby galaxies from
Zaritsky \etal (1994; dashed line).  The grism observations are
well-suited to detect high-redshift sources with very high \OII\ luminosities.  Note here our detection limit (shown in Fig. 12)}
\label{fig:oiilum}
\end{figure}

When we investigate line luminosites of individual knots from the
nearby Zaritsky \etal (1994) sample (Figure~\ref{fig:oiilum}), we see
two distinct distributions, with the high redshift PEARS \OII\
emitters having systematically higher luminosities.  A selection
effect exists here, as we can only detect the brightest sources at
high redshift (detection limit of $log(L(\OII)\/\Lo) \sim 6.3$
in Fig. 15).  The fact that we miss the lower luminosity \OII\
emitters in this dataset does not negate the fact that we \emph{do}
see very high luminosity sources at this redshift in the grism data (see also discussion of high-EW \OII sources in previous section).

\subsection{Galaxies with multiple emitting knots}
Individual HII regions in nearby galaxies have been studied for some
time (Shields 1974; McCall, Rybski, \& Shields 1985, Zaritsky,
Kennicutt, \& Huchra 1994, etc.), and it is well known that star
formation properties vary in local galaxies from one star forming
region to another.  For example, Kennicutt, Edgar, \& Hodge (1989)
find that the luminosity function of HII regions in spiral arms and
interarm regions differ greatly.  Zaritsky, Kennicutt, \& Huchra
(1994) also find measurable differences in line luminosities of
individual star forming regions in nearby galaxies.  Gordon et
al. (2004) investigate in detail the variations of \Ha (as well as UV
and infrared) star formation rates in the many star-forming regions of
M81.  Additionally, star-forming clumps in the interacting system
IC2163 \& NGC2207 are studied by Elmegreen et al. (2006).  However,
investigation of individual emitting regions in high-redshift galaxies has not
been explored as extensively.  Studies of this kind focus on
physical processes occuring within the galaxy, and comparisons of the
high redshift sample to local galaxies help to sort out possible
evolutionary effects.  As noted above, 12 of our 63 2D-selected
emission-line galaxies ($\sim\!20$\%) have multiple emitting knots,
many of which display multiple lines.  These galaxies lie in the
redshift range of $0.12\cle$z$\cle0.44$, the faintest of which has a
continuum magnitude of $i'_{AB}=23.64$ mag.  Here we focus on several
of these PEARS galaxies that have spatially distinct emitting knots.

PEARS Object \# 75753/78237 (SExtractor extracted this object as two
separate objects, but visual inspection shows that the two selected
regions are part of the same galaxy) has six separate emitting
regions, four of which have both \Ha and \OIII\ emission (one knot
containing only \OIII\ emission).  This object is shown in
Figure~\ref{fig:75753}.  In the three knots (which are in 75753) that
have both lines, the \OIII\ flux is approximately 2$\times$ the \Ha
flux (which indicates high excitation, which often means low
metallicity).  The other knot (in 78237) that has both lines has
roughly equal flux in both \Ha and \OIII.  The \Ha flux differs by a
factor of up to $\sim$3.5, suggesting a variation in star formation
rate across the complex structure of this galaxy.  The knot indicated
by the red circle in Figure 11 has an unidentified line with a
wavelength inconsistent with the others present in this source and
with no viable alternative line at this redshift
($\lambda_{obs}$=7872\AA\ ;$\lambda_{RF}$=5875\AA\ if the line originated
from this galaxy).  We also note that this object has a slightly
different color than the rest of the nuclear region of the galaxy.
Given these factors, we conclude that the ``knot'' within the red
circle is an interloper at an undetermined redshift, whose emission
line is present in the spectrum of Object 75753/78237.

PEARS Objects 70314 and 78491 each have three emitting knots (Figures
~\ref{fig:70314} and~\ref{fig:78491} respectively).  Object \# 70314's
three knots all have \Ha emission, with two knots having roughly equal
\Ha flux and equivalent widths, and the other knot having lower flux
and equivalent width values by a factor of $\sim$4.  This galaxy is a
good example of the success of the 2D line-finding method: the lines
in this galaxy---with its multiple blue star-forming regions---were
not detected with the 1D method in either the PEARS or GRAPES data,
because the line flux was washed out by the continuum flux from the
galaxy's core (see also Fig.~\ref{fig:exspec}).  Following this same
line of reasoning, the weakest line of the three is the one
originating from the nucleus of the galaxy \# 70314
(Fig.~\ref{fig:70314}).  The lines stand out more when extracted
solely from the emitting knots instead of the entire galaxy.  All of
the lines associated with Object \# 78491 have equivalent widths
$\gtrsim\!100$~{\AA} (Table 1).  Two of the knots contain both \Ha and
\OIII, and originate from blue star-forming regions on the ends of the
galaxy disk (Fig.~\ref{fig:78491}).  The third ``knot'' associated
with this PEARS ID appears to be another object and has a strong line
($EW\!=\!237.0$~\AA) that remains unidentified due to lack of redshift
for this particular object ($\lambda_{obs}$=7143\AA\
;$\lambda_{RF}$=5788 for $z$=0.234, the redshift of Object \# 78491) .

PEARS Objects 63307, 70407, 75547, 77558, 79283, 79483, 81944, and
88580 all have two emitting knots.  The properties of these galaxies
are given in Table 1.  Object \# 81944 has strong \Ha and \OIII\
emission from one of its knots and a relatively weak \Ha line in the
other (with an equivalent width of $\sim\!4.5$ times lower).  There are
other cases (IDs 79283 and 88580, for example; $z\!=\!0.23$ and $z\!=\!0.269$
respectively) where line flux differs by a factor of 2 or more across
a single galaxy.  This indicates that the star formation properties of
these objects differ across the galaxy itself, and that this effect in
general can be probed at redshifts $z\!\gtrsim\!0.2$.

We also note here that five of the twelve galaxies with multiple emitting
knots did not have detected lines in the deeper GRAPES+1 ACS field
data (described above; Sec. 4.4).  Among the galaxies that were
detected in GRAPES, the PEARS-detected lines' equivalent widths were
higher in every case by on average a factor of$\sim\!4\!\times$.  This
again underscores the strength of the 2D method, which serves to
isolate line emission from individual knots, such that continuum
emission from the rest of the galaxy does not dominate the spectrum.

Given the subset of PEARS HUDF galaxies that exhibit multiple emitting
knots, we expect to have a statistically significant sample of these
objects once analysis of the entire PEARS dataset is completed.  This
should allow an in-depth study of localized star formation at galaxies
up to $z\!\sim\!0.4\!-\!0.5$.

\section{Summary and Future Work}

In summary, it is clear that although each method has some unique
detections, Method 2D-B (triangulation) in general is quite efficient
at detecting emission-line sources in the PEARS grism data, especially
for objects with knotty structures or strong continuum that remain
undetected with the 1D method.  The reason for this advantage is that
the 1D method gives line flux integrated over the whole galaxy, while
the 2D method gives line flux from the emitting region only
(i.e. galaxy knots).  Triangulation requires observations obtained at
multiple roll angle, and hence may not be suitable to all grism
datasets.  Method 2D-A (cross-correlation) can be used with ACS grism
data obtained at one PA (Meurer \etal 2007), but is not fully
automated and may produce false identification of emitting sources in
confused regions such as in extended galaxies.  The triangulation
method will be utilized in future studies of the remaining eight PEARS
fields.  Given the sample of 81 distinct emitting regions, we expect a
total sample of $\sim\!600\!-\!700$ ELGs to continuum
$i'_{AB}\!\lesssim\!26.5$~mag from the combined depth and area of the
PEARS North and South fields.  From this larger statistical sample,
two primary investigations will follow.  First, we will derive line
luminosities, which should allow us to constrain the luminosity
function for \Ha, \OII, and \OIII, going fainter and to higher
redshifts than previous studies.  Secondly, we will use the
luminosities and equivalent widths to estimate the change in the
cosmic star formation rate, again utilizing the depth and quantity of
the PEARS data.  In order to account for dust attenuation, we plan to
estimate an average correction for the different redshift bins in the
survey following previous authors. For example, Davoodi et al. (2007)
calculated estimates for a sample of 1113 SDSS galaxies in the SWIRE
survey (using the extinction law from Calzetti et al. (1994)).  Kewley
et al.  (2004) used the redding curve of Cardelli, Clayton, \& Mathis
(1989).  Additionally, Thompson et al. (2006) give examples of
extinction and surface brightness dimming corrections to star
formation rates for HUDF galaxies with redshifts 1-6.  Pixel-to-pixel
SED decompositions of HUDF galaxies are currently being performed by
Ryan et al. (2008, in prep.) and will also provide estimates for the
effects of dust over a large redshift range.  In addition to these two
primary goals, we will also investigate in further detail the possible
evolution of \OII\ equivalent width and luminosity with redshift
(Figs.~\ref{fig:zvsew} \&~\ref{fig:oiilum}).  The $z=0-1.5$ range is
where the SFR density shows its strongest evolution.  Use of the grism
to isolate the strongest EW sources in this redshift range, combined
with deep HST imaging, will prove to be an excellent way to select
galaxies that are most evolving over this important redshift range.
We will then perform a detailed quantitative study of the morphologies
of these objects, so as to diagnose what is causing the evolution.
Simulations of the PEARS data (which are currently being performed)
will allow us to gain a better insight into various selection effects
and limits, and will aid in conclusions drawn from the dataset.  These
future studies should provide a more detailed look at the overall
nature of these line-emitting galaxies, thus revealing mechanisms of
star forming activity over $z\!=\!0\!-\!1.5$.

This research was supported in part by the NASA Harriett G. Jenkins Predoctoral Fellowship (ANS), and by grants HST-GO-10530 \& HST-GO-9793 from STScI, which is
operated by AURA for NASA under contract NAS 5-26555.  We thank the anonomous referee for helpful comments which improved the paper.

\begin{deluxetable}{cccccccccc}
\tabletypesize{\scriptsize}
\tablecaption{Global Properties of Emission-Line Galaxies \label{table1}}
\tablewidth{0pt}
\tablehead{
\colhead{PEARS}&\colhead{Knot}&\colhead{RA}&\colhead{Dec}&\colhead{$i'_{AB}$}&\colhead{Wavelength}&\colhead{Flux}&\colhead{Equivalent Width}&\colhead{Line}&\colhead{Grism}\\
\colhead{ID}&\colhead{\# }&\colhead{(deg)}&\colhead{(deg)}&\colhead{(mag)}&\colhead{(\AA)}&\colhead{($10^{-18} erg/s/cm^2$)}&\colhead{(\AA)}&\colhead{ID}&\colhead{Redshift}
}
\startdata
63307 & 4 & 53.1433945 & -27.8134537 & 18.65 & 7938 & 8.8$\pm$4.8 & \nodata & \Hbeta & 0.633 \\ 
63307 & 5 & 53.1441269 & -27.8134251 & 18.65 & 8258 & 12.8$\pm$7.9 & \nodata & \OIII & 0.653 \\ 
68739 & 1 & 53.1607895 & -27.8163128 & 24.88 & 7570 & 9.3$\pm$2.1 & 125.6 & \OIII & 0.516 \\ 
70314 & 1 & 53.1748161 & -27.7995949 & 20.18 & 7525 & 28.4$\pm$13.9 & 68.9 & \Ha & 0.147 \\ 
70314 & 2 & 53.1750183 & -27.7993336 & 20.18 & 7527 & 29.1$\pm$5.6 & 86.3 & \Ha & 0.147 \\ 
70314 & 3 & 53.1747475 & -27.7992420 & 20.18 & 7455 & 6.6$\pm$4.9 & 20.4 & \Ha & 0.137 \\ 
70407 & 1 & 53.1851540 & -27.8052826 & 20.41 & 9300 & 60.4$\pm$26.1 & 28.1 & \Ha & 0.417 \\ 
70407 & 7 & 53.1851730 & -27.8052406 & 20.41 & 9452 & 42.4$\pm$20.0 & 20.2 & \Ha & 0.440 \\ 
70651 & 1 & 53.1530495 & -27.8121529 & 23.32 & 6039 & 80.0$\pm$9.8 & 559.0 & \OIII & 0.209 \\ 
70651 & 1 & 53.1530495 & -27.8121529 & 23.32 & 7956 & 30.4$\pm$3.3 & 295.6 & \Ha & 0.212 \\ 
71864 & 1 & 53.1506119 & -27.8095131 & 24.65 & 8785 & 19.7$\pm$5.2 & 130.8 & \OIII & 0.759 \\ 
71924 & 1 & 53.1536827 & -27.8088989 & 23.84 & 6898 & 7.5$\pm$3.0 & 31.3 & \OIII & 0.381 \\ 
72179 & 1 & 53.1310501 & -27.8084450 & 23.29 & 6281 & 29.1$\pm$10.5 & 93.7 & NA & \nodata \\ 
72509 & 1 & 53.1705208 & -27.8066082 & 24.46 & 8547 & 7.1$\pm$3.1 & 28.1 & \OII & 1.294 \\ 
72557 & 1 & 53.1338768 & -27.8068733 & \nodata & 6677 & 15.7$\pm$3.2 & 477.2 & NA & \nodata \\ 
73619 & 1 & 53.1844063 & -27.8051853 & 24.77 & 8249 & 9.2$\pm$3.1 & 65.1 & \OIII & 0.652 \\ 
74234 & 1 & 53.1377335 & -27.8042202 & 25.90 & 7704 & 10.2$\pm$2.1 & 131.3 & \OIII & 0.542 \\ 
74670 & 1 & 53.1616173 & -27.8027954 & 24.03 & 6543 & 14.9$\pm$7.3 & \nodata & \OIII & 0.310 \\ 
75506 & 1 & 53.1472664 & -27.8008537 & 26.32 & 8418 & 17.1$\pm$7.3 & 1157.5 & \Ha & 0.283 \\ 
75506 & 1 & 53.1472664 & -27.8008537 & 26.32 & 6379 & 16.4$\pm$3.9 & 323.2 & \OIII & 0.277 \\ 
75547 & 1 & 53.1733017 & -27.7993031 & 23.64 & 7372 & 7.8$\pm$2.4 & 59.9 & \Ha & 0.123 \\ 
75547 & 2 & 53.1732941 & -27.7992783 & 23.64 & 7372 & 6.3$\pm$1.9 & 45.8 & \Ha & 0.123 \\ 
75753 & 1 & 53.1872597 & -27.7943401 & 21.57 & 6708 & 67.4$\pm$4.8 & 260.3 & \OIII & 0.343 \\ 
75753 & 1 & 53.1872597 & -27.7943401 & 21.57 & 8816 & 32.8$\pm$4.2 & 179.7 & \Ha & 0.343 \\ 
75753 & 2 & 53.1873703 & -27.7942238 & 21.57 & 6685 & 122.1$\pm$5.9 & 285.0 & \OIII & 0.338 \\ 
75753 & 2 & 53.1873703 & -27.7942238 & 21.57 & 8819 & 65.5$\pm$4.9 & 207.8 & \Ha & 0.344 \\ 
75753 & 3 & 53.1878090 & -27.7939053 & 21.57 & 8800 & 80.0$\pm$6.9 & 123.5 & \Ha & 0.341 \\ 
75753 & 3 & 53.1878090 & -27.7939053 & 21.57 & 6697 & 148.7$\pm$6.6 & 272.3 & \OIII & 0.324 \\ 
76154 & 1 & 53.1512299 & -27.7987995 & 23.67 & 8016 & 37.5$\pm$3.0 & 285.5 & \OIII & 0.600 \\ 
77558 & 8 & 53.1864052 & -27.7910328 & 18.66 & 7995 & 23.1$\pm$5.2 & \nodata & \Ha & 0.218 \\ 
77558 & 11 & 53.1871910 & -27.7909679 & 18.66 & 7890 & 84.6$\pm$15.1 & \nodata & NA & \nodata \\ 
77902 & 1 & 53.1559830 & -27.7949619 & 23.48 & 7720 & 7.4$\pm$2.3 & 53.3 & \OII & 1.071 \\ 
78021 & 1 & 53.1839218 & -27.7954350 & 27.45 & 8614 & 13.3$\pm$3.6 & 294.2 & \OII & 1.311 \\ 
78077 & 1 & 53.1841545 & -27.7926388 & 21.67 & 6482 & 48.2$\pm$17.7 & 32.0 & \OII & 0.739 \\ 
78077 & 1 & 53.1841545 & -27.7926388 & 21.67 & 8675 & \nodata & \nodata & \OIII & 0.737 \\ 
78237 & 1 & 53.1876869 & -27.7943954 & 22.21 & 6693 & 39.8$\pm$6.8 & 166.7 & \OIII & 0.340 \\ 
78237 & 1 & 53.1876869 & -27.7943954 & 22.21 & 8800 & 38.7$\pm$6.8 & 160.4 & \Ha & 0.340 \\ 
78237 & 2 & 53.1879768 & -27.7943249 & 22.21 & 6701 & 70.2$\pm$16.6 & 381.6 & \OIII & \nodata\\ 
78237 & 3 & 53.1877136 & -27.7942200 & 22.21 & 7872 & 20.6$\pm$3.8 & 39.7 & NA & \nodata \\ 
78491 & 1 & 53.1548195 & -27.7934532 & 22.64 & 7143 & 15.4$\pm$3.9 & 237.0 & NA & \nodata \\ 
78491 & 3 & 53.1544189 & -27.7933788 & 22.64 & 8100 & 10.3$\pm$2.7 & 224.3 & \Ha & 0.234 \\ 
78491 & 3 & 53.1544189 & -27.7933788 & 22.64 & 6120 & 17.2$\pm$3.8 & 190.1 & \OIII & 0.234 \\ 
78491 & 4 & 53.1547127 & -27.7931709 & 22.64 & 6114 & 54.5$\pm$5.4 & 275.6 & \OIII & 0.234 \\ 
78491 & 4 & 53.1547127 & -27.7931709 & 22.64 & 8080 & 20.4$\pm$3.0 & 144.1 & \Ha & 0.234 \\ 
78582 & 2 & 53.1615829 & -27.7922630 & 21.16 & 9541 & 276.2$\pm$54.4 & 189.4 & \Ha & 0.454 \\ 
78582 & 2 & 53.1615829 & -27.7922630 & 21.16 & 7100 & \nodata & \nodata & \OIII & 0.454 \\ 
78762 & 1 & 53.1618958 & -27.7925568 & 22.79 & 7283 & 28.8$\pm$3.9 & \nodata & \OIII & 0.458 \\ 
79283 & 2 & 53.1419983 & -27.7867641 & 20.75 & 8070 & 19.3$\pm$5.9 & 36.2 & \Ha & 0.230 \\ 
79283 & 3 & 53.1421967 & -27.7865429 & 20.75 & 8059 & 37.7$\pm$5.9 & \nodata & \Ha & 0.230 \\ 
79400 & 1 & 53.1673317 & -27.7918015 & 23.92 & 6866 & 12.1$\pm$4.7 & 47.8 & \OIII & 0.375 \\ 
79483 & 1 & 53.1879654 & -27.7900734 & 20.80 & 9200 & 94.9$\pm$17.6 & 54.0 & \Ha & 0.483 \\ 
79483 & 2 & 53.1879539 & -27.7900009 & 20.80 & 9437 & 130.6$\pm$23.6 & 55.5 & \Ha & 0.483 \\ 
79483 & 2 & 53.1879539 & -27.7900009 & 20.80 & 7001 & 14.7$\pm$8.9 & \nodata & \OIII & 0.483 \\ 
79520 & 1 & 53.1861954 & -27.7916622 & 23.78 & 8703 & 12.5$\pm$2.9 & 73.1 & \OIII & 0.742 \\ 
80071 & 1 & 53.1866226 & -27.7902203 & 23.55 & 7334 & 47.2$\pm$3.0 & 224.1 & \Ha & 0.118 \\ 
80255 & 1 & 53.1848145 & -27.7899342 & 23.62 & 7277 & 5.5$\pm$1.6 & \nodata & \OII & 0.953 \\ 
80500 & 1 & 53.1472092 & -27.7884693 & 23.34 & 8300 & 15.3$\pm$3.4 & 60.8 & \OIII & 0.658 \\ 
80500 & 1 & 53.1472092 & -27.7884693 & 23.34 & 6178 & 11.8$\pm$3.5 & 42.9 & \OII & 0.658 \\ 
80666 & 1 & 53.1765137 & -27.7897243 & 24.96 & 7047 & 27.6$\pm$3.7 & 280.7 & \OIII & 0.411 \\ 
81032 & 1 & 53.1815071 & -27.7879314 & 23.33 & 6044 & 34.1$\pm$9.8 & 119.2 & \OIII & 0.210 \\ 
81256 & 1 & 53.1920815 & -27.7872849 & 23.02 & 7840 & 7.9$\pm$2.4 & 36.4 & \OII & 1.104 \\ 
81609 & 1 & 53.1640930 & -27.7872963 & 24.35 & 7820 & 15.7$\pm$2.8 & 60.7 & \OII & 1.098 \\ 
81944 & 1 & 53.1446838 & -27.7855377 & 22.48 & 8138 & 62.1$\pm$4.5 & 74.8 & \Ha & 0.228 \\ 
81944 & 2 & 53.1447372 & -27.7854137 & 22.48 & 6132 & 401.0$\pm$12.5 & 483.2 & \OIII & 0.228 \\ 
81944 & 2 & 53.1447372 & -27.7854137 & 22.48 & 8129 & 162.1$\pm$9.4 & 341.4 & \Ha & 0.228 \\ 
82307 & 1 & 53.1634598 & -27.7866497 & 25.25 & 7369 & 15.5$\pm$2.9 & \nodata & \OIII & 0.475 \\ 
83381 & 1 & 53.1765251 & -27.7825947 & 24.91 & 6640 & 24.7$\pm$2.8 & 124.6 & \OIII & 0.329 \\ 
83553 & 1 & 53.1784821 & -27.7840424 & 24.82 & 6461 & 94.9$\pm$5.2 & 93.2 & \CIV & 3.166 \\ 
83553 & 1 & 53.1784821 & -27.7840424 & 24.82 & 7940 & 13.3$\pm$3.3 & 26.8 & \CIII & 3.166 \\ 
83686 & 1 & 53.1518135 & -27.7829018 & 23.44 & 6847 & 8.9$\pm$3.1 & 39.0 & \OII & 0.837 \\ 

\enddata
\tablenotetext{*}{NOTE: NA indicates line could not be identified. No data indicates measurement was not possible  ``Grism Redshift'' column gives re-calculated redshift based on the line identification.}
\end{deluxetable}

\begin{deluxetable}{cccccccccc}
\tabletypesize{\scriptsize}
\tablecaption{Global Properties of Emission-Line Galaxies cont.}
\tablewidth{0pt}
\tablehead{
\colhead{PEARS}&\colhead{Knot}&\colhead{RA}&\colhead{Dec}&\colhead{$i'_{AB}$}&\colhead{Wavelength}&\colhead{Flux}&\colhead{Equivalent Width}&\colhead{Line}&\colhead{Grism}\\
\colhead{ID}&\colhead{\# }&\colhead{(deg)}&\colhead{(deg)}&\colhead{(mag)}&\colhead{(\AA)}&\colhead{($10^{-18} erg/s/cm^2$)}&\colhead{(\AA)}&\colhead{ID}&\colhead{Redshift}
}
\startdata
83789 & 1 & 53.1527901 & -27.7826843 & 24.74 & 8861 & 31.2$\pm$10.7 & 229.4 & \OIII & 0.774 \\ 
83804 & 1 & 53.1845818 & -27.7833576 & 24.96 & 7918 & 9.0$\pm$2.6 & 70.0 & \OII & 1.125 \\ 
83834 & 1 & 53.1580925 & -27.7812119 & 21.94 & 8101 & 8.1$\pm$2.4 & 53.3 & NA & \nodata \\ 
85517 & 1 & 53.1763344 & -27.7808685 & 24.79 & 7644 & 25.0$\pm$3.5 & 205.2 & \OIII & 0.530 \\ 
85844 & 1 & 53.1624680 & -27.7803612 & 25.77 & 8568 & 22.4$\pm$3.8 & 240.7 & \OII & 1.299 \\ 
87294 & 1 & 53.1629181 & -27.7752514 & 21.02 & 8191 & 8.0$\pm$2.1 & 194.1 & NA & \nodata \\ 
87464 & 1 & 53.1878166 & -27.7726479 & 22.53 & 7460 & 61.7$\pm$9.2 & 170.9 & \Ha & 0.130 \\ 
87464 & 1 & 53.1878166 & -27.7726479 & 22.53 & 5642 & 123.1$\pm$20.7 & 394.7 & \OIII & 0.130 \\ 
87658 & 1 & 53.1477661 & -27.7769241 & 23.96 & 7853 & 7.8$\pm$2.4 & 30.6 & \OII & 1.107 \\ 
88580 & 1 & 53.1620064 & -27.7740345 & 22.60 & 6354 & 12.8$\pm$2.9 & 75.0 & \OIII & 0.269 \\ 
88580 & 2 & 53.1619110 & -27.7738514 & 22.60 & 6338 & 30.6$\pm$4.5 & 96.6 & \OIII & 0.269 \\ 
89030 & 1 & 53.1604347 & -27.7752380 & 25.79 & 9126 & 32.3$\pm$4.3 & 101.3 & \OII & \nodata \\ 
89209 & 1 & 53.1503944 & -27.7720318 & \nodata & 6075 & 26.2$\pm$4.7 & 182.0 & \OIII & 0.216 \\ 
89209 & 1 & 53.1503944 & -27.7720318 & \nodata & 8000 & 26.2$\pm$4.7 & 182.0 & \Ha & 0.216 \\ 
89853 & 1 & 53.1375847 & -27.7691345 & 21.62 & 8951 & 17.3$\pm$5.4 & 20.1 & \Ha & 0.364 \\ 
89923 & 1 & 53.1739769 & -27.7720718 & 21.24 & 8750 & 37.8$\pm$6.9 & 43.5 & \Ha & 0.333 \\ 
90116 & 1 & 53.1948090 & -27.7733440 & 25.44 & 8143 & 23.0$\pm$3.5 & 325.2 & \OIII & 0.630 \\ 
90246 & 1 & 53.1512070 & -27.7728481 & 24.03 & 8008 & 20.7$\pm$3.4 & 110.8 & \OIII & 0.603 \\ 
91205 & 1 & 53.1505280 & -27.7713089 & 23.17 & 7856 & 26.6$\pm$7.5 & 73.1 & \Ha & 0.197 \\ 
91789 & 1 & 53.1470642 & -27.7701302 & 23.80 & 7655 & 11.8$\pm$3.2 & 66.4 & \OIII & 0.533 \\ 
92839 & 1 & 53.1628647 & -27.7671719 & 20.95 & 6201 & \nodata & \nodata & MgII & 1.215 \\ 
94632 & 1 & 53.1795502 & -27.7662010 & 24.93 & 8309 & 11.6$\pm$2.7 & 82.9 & \OIII & 0.664 \\ 
95471 & 1 & 53.1773453 & -27.7639313 & 22.37 & 8001 & 23.2$\pm$8.4 & 51.2 & \Ha & 0.219 \\ 
96123 & 1 & 53.1429062 & -27.7636814 & 23.11 & 7664 & 9.1$\pm$2.7 & 22.3 & \OIII & 0.535 \\ 
96627 & 1 & 53.1704559 & -27.7614193 & 21.50 & 7453 & 30.7$\pm$8.7 & 51.8 & \Ha & 0.136 \\

\enddata
\tablenotetext{*}{NOTE: NA indicates line could not be identified. No data indicates measurement was not possible  ``Grism Redshift'' column gives re-calculated redshift based on the line identification.}
\end{deluxetable}


\end{document}